\documentclass[preprints,article,accept,moreauthors,pdftex]{mdpi} 

\firstpage{1} 
\pubvolume{xx}
\issuenum{1}
\articlenumber{5}
\pubyear{2019}
\copyrightyear{2019}
\history{Received: date; Accepted: date; Published: date}

\Title{Towards Highly Efficient Polymer Fiber Laser Sources for Integrated Photonic Sensors}

\Author{Simon Spelthann $^{1,2}$\orcidA{}, Stefanie Unland$^{2}$, Jonas Thiem$^{1,2}$\orcidB{}, Florian Jakobs$^{3}$, Jana Kielhorn$^{3}$, Pen Yiao Ang$^{3}$, Hans-Hermann Johannes$^{3,4,5}$, Dietmar Kracht$^{2,5}$, Joerg Neumann$^{2,5}$, Axel Ruehl$^{1,2,4}$*, Wolfgang Kowalsky$^{3,4,5}$, and Detlev Ristau $^{1,2,4,5}$}

\AuthorNames{Simon Spelthann, Stefanie Unland, Jonas Thiem, Florian Jakobs, Jana Kielhorn, Pen Yiao Ang, Dietmar Kracht, Joerg Neumann, Hans-Hermann Johannes, Axel Ruehl, Wolfgang Kowalsky, and Detlev Ristau}

\address{%
$^{1}$ \quad Leibniz University Hannover, Institute of Quantum Optics\\
$^{2}$ \quad Laser Zentrum Hannover e.V.\\
$^{3}$ \quad TU Braunschweig, Institut für Hochfrequenztechnik\\
$^{4}$ \quad Academic Alliance Braunschweig -- Hannover QUANOMET\\
$^{5}$ \quad Cluster of Excellence PhoenixD}

\corres{Correspondence: ruehl@iqo.uni-hannover.de}

\abstract{
	Lab-on-a-Chip (LoC) devices combining microfluidic analyte provision with integrated optical analysis are highly desirable for several applications in biological oder medical sciences. While the microfluidic approach is already broadly adressed, yet some work needs to be done regarding the integrated optics, especially provision of highly integrable laser sources.
	Polymer optical fiber (POF) lasers represent an alignment-free, rugged and flexible technology platform. Additionally, POFs are intrinsically compatible to polymer microfluidic devices. Home-made Rhodamine B (RB) doped POFs were characterized with experimental and numerical parameter studies on their lasing potential. 
	High output energies of $1.65\,mJ$, high slope efficiencies of $56\,\%$ and $50\,\%$-lifetimes of $\leq  900\,k$ shots were extracted from RB:POFs. Furthermore, RB:POFs show broad spectral tunability over several tens of nanometers.
	A route to optimize polymer fiber lasers is revealed providing functionality for a broad range of LoC devices. Spectral tunability,  high efficiencies and output energies enable a broad field of LoC applications. 
}


\begin{document}

\section{Introduction}
Sensors for biological or medical sciences are often bulky and complex. Especially, optical analysis techniques usually consist of serveral distinct devices like lasers, optical components and the analysis platform. As a consequence, such devices are not only complex to operate and maintain but also expensive. However, for spectroscopic approaches e.g. in point-of-care testing it is highly desirable to provide cheap, small, and easy to use and fabricate devices. 

These requirements have been approached by Lab-on-a-Chip (LoC) devices, e.g. miniaturized gas chromatography systems holding promising potential for the rapid analysis on a compact and fully integrable platform \cite{akbar, tao, paknahad}.
Combining integrated photonics such as laser systems with existing LoC approaches is mandatory to fulfill the requirements for modern optical analysis techniques \cite{zhou, tao}. Such LoC devices seem to be in the scope as 3d printers have become more and more sophisticated and affordable \cite{vaezi, ngo}. Polymers are the easiest to handle via 3d printing and thus constitute one of the most promising laser platforms for LoC devices. Semiconductor light sources e.g. laser diodes (LD) or light emitting diodes (LED) indeed combine a compact design with tailored emission properties and all-electrically control but the integration into a polymer platform is difficult and usually done by means of assembly. One batch producable LoC devices would lead to a higher cost effectiveness requiring polymerizable lightsources. Organic LEDs were combined with organic photodiodes on glass substrates and then attached to a microfluidic channel for fluorescent sensing \cite{shu}. Recently, electrically pumped organic LDs were reported for the first time emitting an optical power of $0.5\, mW $ \cite{sandanayaka}. Although this approach seems to be promising, the high attenuation in polymers of up to several $dB/cm$ requires higher power levels. 

In this paper, we adress the needs for laseractive gain materials capable to produce high output energies while being easily integrable into a polymer platform: polymer fiber lasers. They provide a rugged and flexible technology platform which enables simple, compact and alignment-free laser setups. Fiber lasers based on polymers meet the requirement for compatibilty with microfluidic platforms on the one hand and 3d printing processes on the other hand. Polymer fiber lasers thus constitute an excellent substitute light source for polymer based LoC devices.

Laser activity in polymer optical fibers (POF) are mostly realized by doping the fiber core with laser dyes. Highly efficient lasers can be set up performing on slope efficiencies up to $43\,\%$ and providing high output energies of $640\,\mu J$ \cite{kuriki1}. Large gain values of up to $27\,dB$ can be exploited in fiber amplifiers \cite{tagaya}, but degradation of the laser dyes via photobleaching and thermal load during the pumping process are general weak points resulting in $50\,\%$-lifetimes of $200\,k$ pulses at repetition rates of $10\,Hz$ \cite{kuriki2}. 
Hence, the dye molecules can be damaged when treated thermally. Regarding 3d printing processing polymers above $200\,^\circ C$, the efficiency of a laser will be reduced when integrating the active material in a photonic device. To nevertheless provide sufficient optical output energies, the performance of RB:POF lasers has to be optimized.

We present experimental and numerical studies on the optimization of POF lasers using polymethyl methacrylate (PMMA) doped with Rhodamine B (RB). The lasing characteristics such as wavelength and slope efficiency were investigated for laser output couplers (OC) with different degree of reflection. Laser experiments were compared to numerical simulations.
 
\section{Manufacturing and Characteristics}\label{sec:manufacturing}
The PMMA preforms were prepared using radical polymerization in bulk with $0.03 \,mol\%$ lauroyl peroxide as initiator and $0.1\, mol\%$ n-butyl-mercaptan as chain transfer agent. RB was mixed with the initiator and the chain transfer agent in nitrogen-saturated methyl methacrylate. The solution was polymerized and the final RB:PMMA preforms were drawn in a self-made drawing tower to fiber cores. In a second step the fiber is coated with acrylate and drawn through a nozzle. The cladding was cured with UV light afterwards. The diameter of the resulting core ranged from $920-1000\,\mu m$, whereas the total diameter was $1150\,\mu m$. More details on the preparation and characterization process for dye doped POF can be found in \cite{caspary, zaremba}. Fibers with different doping concentrations of $1\,ppm$, $5\,ppm$ and $10\,ppm$ were manufactured and used for the experiments. Note that these values refer to the amount of RB which was added to the synthesis. The exact value may slightly deviate due to the thermal treatment during the polymerization and the fiber drawing process. 

Absorption and emission spectra, fluorescence lifetime as well as the attenuation of the fiber were measured to deduce experimental parameters but also for use as input parameters for numerical investigations concerning the RB:POF lasers. 

\begin{figure}[H]
	\centering
	\includegraphics[width=15.5 cm]{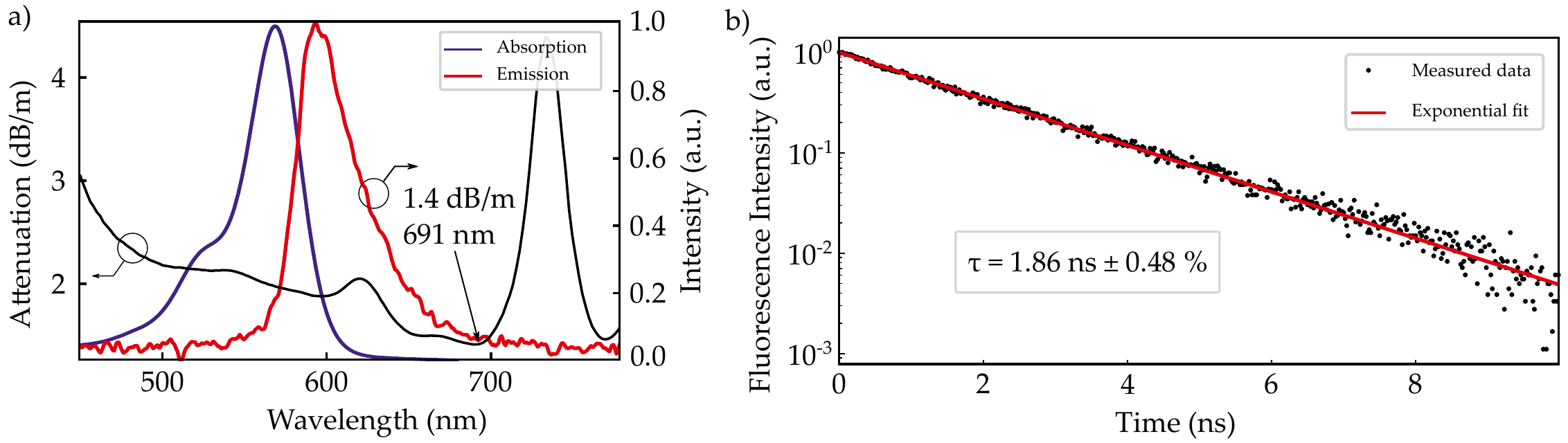}
	\caption{Spectral characteristics of the used RB:POFs. a) Absorption and Emission spectra together with the spectral attenuation. b) Fluorescence lifetime of RB doped PMMA used for fiber manufacturing.}
	\label{fig:spectral}
\end{figure}

The absorption spectrum was measured for RB doped PMMA preforms with a doping concentration of $1\,ppm$ with the maximum located at $557\,nm$ as depicted in Fig. \ref{fig:spectral} a). The absorption maxima of $5\,ppm$ and $10\,ppm$ differ by $\Delta \lambda \leq  2\,nm $ consistent with previous reports, e.g. Ref. \cite{ahmed}. The emission spectra are red shifted with respect to the absorption spectra due to the Stokes shift. Depending on the doping concentration the emission maxima range between $595\,nm$ and $618\,nm$. The absorption and emission spectra overlap which causes reabsorption resulting in a greater Stokes shift for high doping concentrations \cite{bavali}. We further manufactured a passive POF in the same way as described above. Cut back experiments were carried out with a supercontinuum light source to determine the spectral dependence of the fiber attenuation depicted in Fig. \ref{fig:spectral} a). Infrared attenuation is not shown in the figure because the attenuation exceeds values which are reasonable for laser applications. The attenuation minimum of $1.4\,dB/m$ is located at $691\,nm$. Measurements of the fluorescence decay with a streak camera resulted in a fluorescence lifetime of $1.86\,ns$ as shown in Fig. \ref{fig:spectral} b). Such low upper state lifetimes result in high lasing thresholds thus implying high pump energies e.g. available from q-switch pulsed laser systems. 
\section{Numerical Model}\label{sec:numerical}
We used a system of rate equations to verify our experimental results numerically. These rate equations were discretised for spatial dimension $z$ and the wavelength $\lambda $ of the propagating light. Steady-state operation of our RB:POF laser can be modelled with a system of two energy levels with population numbers $N_2$ of the upper laser level and $N_1$ of the lower laser level, where the derivatives with respect to time $d/dt$ are set to zero \cite{arrue}: %
\begin{equation}
\frac{d N_2}{dt} = -\frac{N_2}{\tau} + \left( \frac{\sigma_{abs}(\lambda) P^{\pm}(z,\lambda)}{h\nu A}N_1- \frac{\sigma_{em}(\lambda) P^{\pm}(z,\lambda)}{h\nu A}N_2 \right)
\label{eq:rateeq1}
\end{equation}
\begin{equation}
\frac{d N_1}{dt} = - \frac{\sigma_{abs}(\lambda) P^{\pm}(z,\lambda)}{h\nu A}N_1 + \frac{\sigma_{em}(\lambda) P^{\pm}(z,\lambda)}{h\nu A}N_2 
\label{eq:rateeq2}
\end{equation}
\begin{equation}
N_{total}=N_1+N_2
\label{eq:rateeq3}
\end{equation}

In these equations, $ \tau $ is the fluorescence lifetime of the upper laser level 2, $\sigma_{abs}$ and $\sigma_{em} $ are the cross sections for absorption and emission, respectively. $N_1$ and $N_2$ constitute the population of the corresponding energy levels and $A$ is the cross section of the fiber core. Forwards ($+$) and backwards ($-$) propagation are superscripted in the equations. All parameters were set to SI units. The propagation of the laser power can be described by 

\begin{equation}
\frac{dP^{\pm}}{\pm dz} = \sigma_{em}(\lambda) P(z, \lambda)^{\pm} N_2 - \sigma_{abs}(\lambda) P(z, \lambda)^{\pm} N_1 + \frac{N_2 h c}{\tau \lambda} \epsilon_{em, sp} (\lambda) \beta A
\label{eq:propagation}
\end{equation}
where $\beta$ is the amount of fluorescence that is guided within the fiber which depends on the numerical aperture of the fiber. The efficiency of spontaneous emission $ \epsilon_{em, sp} $ is calculated by normalizing the corresponding cross section $\sigma_{em}$ to an integral value of 1.  We used the shooting method \cite{sujecki} to numerically solve the equations for the boundary conditions of a fiber laser system with reflectivities in the laser resonator of $R_1$ and $R_2$.%
\begin{equation}
P^-(z, \lambda) = P^+ (l, \lambda ) R_2(\lambda)
\label{eq:boundaryR1}
\end{equation}
with $l$ the length of a particular fiber and
\begin{equation}
P^+(0, \lambda) = P^- (0, \lambda ) R_1(\lambda) \text{.}
\label{eq:boundaryR2}
\end{equation}
The shooting algorithm calculates the propagating Powers $P^+$ and $P^-$ and iterativly minimizes the difference between $P^+(z) \cdot R2 $ and $P^-(z)$ by changing the input power levels at $z=0$.
The pump rate has to be modified as we used a transversal pump scheme. The pump power is constant within each spatial increment $dz$ over the length of the fiber. Transversal pump light distribution is averaged through the fiber cross section. The extracted laser power is calculated from the boundary conditions, where absorption inside the mirror is neglected.
\begin{equation}
P_{Lasing}(\lambda) = P^+(z, \lambda)(1-R_2(\lambda))
\label{eq:lasing_power}
\end{equation}
The spectroscopic data and fiber properties presented in Sections \ref{sec:manufacturing} were used as input parameters to solve equations (\ref{eq:rateeq1}-\ref{eq:lasing_power}). 

\section{RB:POF Laser}\label{sec:laser}
\begin{figure}[H]
	\centering
	\includegraphics[width=7.5 cm]{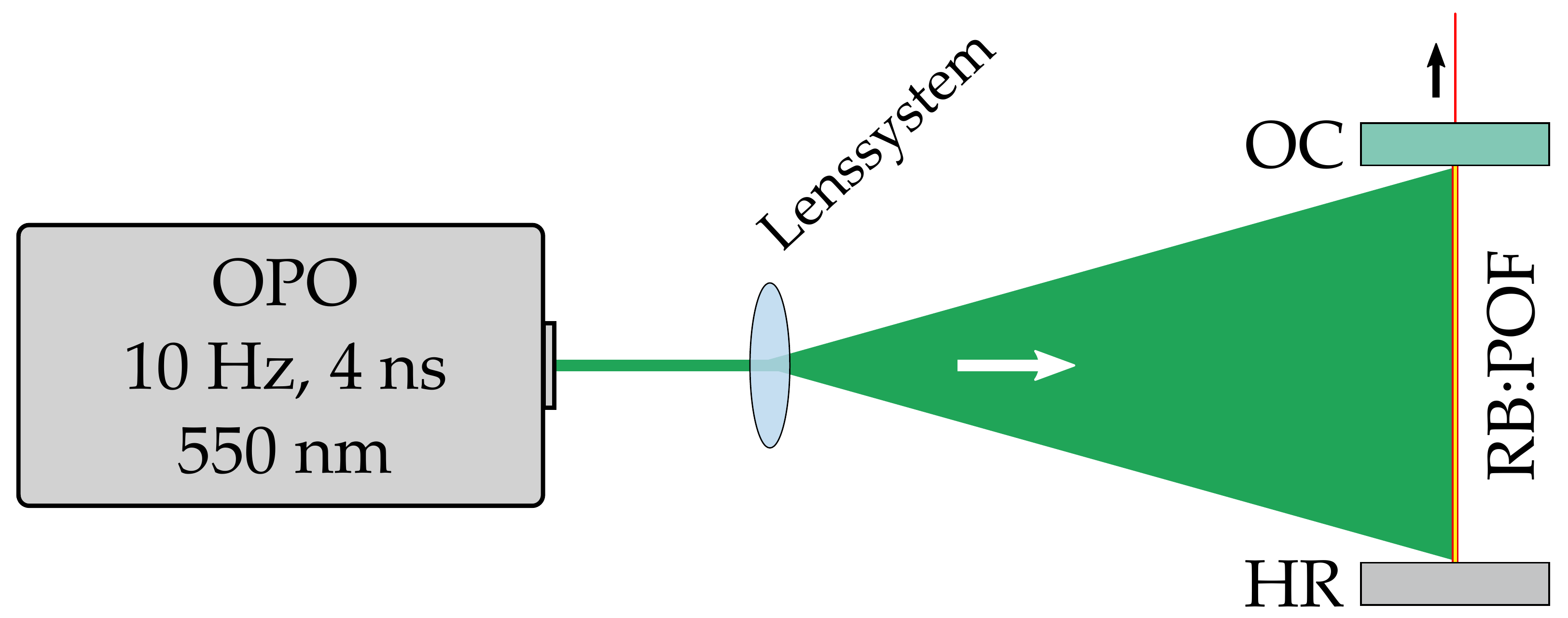}
	\caption{Schematic of the RB:POF laser transversally pumped with an optical parametric oscillator (OPO).The resonator mirrors are both butt coupled to the fiber facets. OC: output coupler; HR: high reflective mirror.}
	\label{fig:aufbauab}
\end{figure}

Laser experiments were performed with a transversal pumping scheme as sketched in Fig. \ref{fig:aufbauab} to avoid bleaching. Photo bleaching may occur due to triplet-triplet transitions within a dye molecule but high pump energies can also destroy the dye molecules thermally \cite{george, kurian}. In preliminary experiments on different pumping schemes in RB:POF amplifiers, the bleaching rate was almost 60-fold lower for the transversal geometry compared to the traditional longitudinal pumping. 

As a pump source we used an optical parametric oscillator (OPO) itself pumped by the third harmonic of a q-switched Nd:YAG laser providing pulse energies up to $70\,mJ$ and pulse durations of $4\,ns$. For optimum pump absorption in our material, the OPO was tuned to $550\,nm$. This deviation from the wavelength of maximum absorption stated in Section \ref{sec:manufacturing} was necessary due to technical reasons. Although the pump wavelength set in our experiments deviates by $7\,nm$, pump absorption is estimated to differ by only $1.5\,\%$ which was considered to be acceptable. The laser consisted of a linear cavity including a $6\,cm$ gain fiber. In order to prepare the RB:POFs for the experiments, they were cut into the particular length using a POF-cutter. To release internal stress originating from the fiber drawing process the fibers were tempered in an oven at $95\,^\circ C$ for $10 \, min$. Afterwards, the facets were polished in a three-step procedure to obtain a clean and smooth facet surface. The fiber was mounted in an SMA plug connected to a polishing puck. Polishing films with grain sizes of $3\,\mu m$, $1\,\mu m$ and $0.3\,\mu m$ were used to polish the fibers on a glass plate.

A silver coated high reflective mirror and a variable dielectric output coupling mirror (OC) both butt-coupled to the fiber endfaces formed the resonator. Note, that the wavelength dependency of the used OC reflectivity results in small deviations of the reflectivity for the differently doped fibers. Different doping concentrations as well as the OC reflectivity were investigated in a parameter study to determine optimum parameters towards high efficiencies and output energies.

\subsection{Optimization of the Laser Efficiency}\label{subsec:laseff}
\begin{figure}[H]
	\centering
	\includegraphics[width=16 cm]{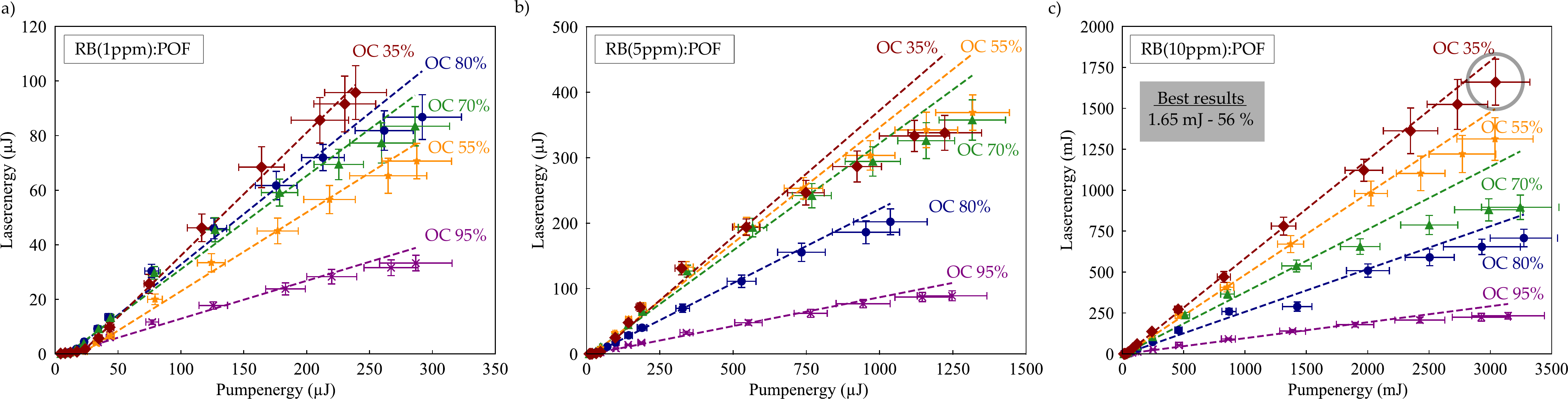}
	\caption{Characteristic curves of the lasers for doping concentrations of a) $1\,ppm$, b) $5\,ppm$ and c) $10\,ppm$. Saturation effects were observed, especially for the $5\,ppm$ doped fiber laser.}
	\label{fig:efficiency}
\end{figure}

Output versus pump characteristics of the lasers were recorded and are depicted in Fig. \ref{fig:efficiency}. Accurate values of the absorbed pump power were obtained experimentally by measuring the transversally transmitted pump energy through an aperture. The slope efficiencies were obtained from a linear fit. In general, laser output energy increases with decreasing OC reflectivity as high amplifications can be exploited from dye doped POFs. The laser characteristics for the $1\,ppm$ deviate from this observation which might be due to nonuniform distribution of RB molecules within our preforms. Saturation effects were observed for doping concentrations $>5\,ppm$. Increasing doping concentrations only led to increasing output energies as long as reabsorption and bleaching play a minor role. 
We observed a maximum pulse energy of $1.65\,mJ$ by using a $10\,ppm$ POF together with a $35\,\%$ output coupler. Taking into account the pulse durations of about $2\,ns$ mesured with a fast photodiode, this corresponds to a peak power of $825\,kW$ and constitutes an almost three-fold improvement to the RB:POF laser presented by Kuriki et al. \cite{kuriki1}. In this configuration, the slope efficiency of $56\,\%$ was the maximum value in the parameter range used in this study. 
This constitutes an improvement of more than 10 percentage points to the RB:POF laser of \cite{kuriki1}.
To the best of our knowledge, our $10\,ppm$ doped RB:POF laser with an OC reflectivity of $35\,\%$ thus performed on the highest slope efficiency and provided the highest output energy. 

\begin{figure}[H]
	\centering
	\includegraphics[width=7.5 cm]{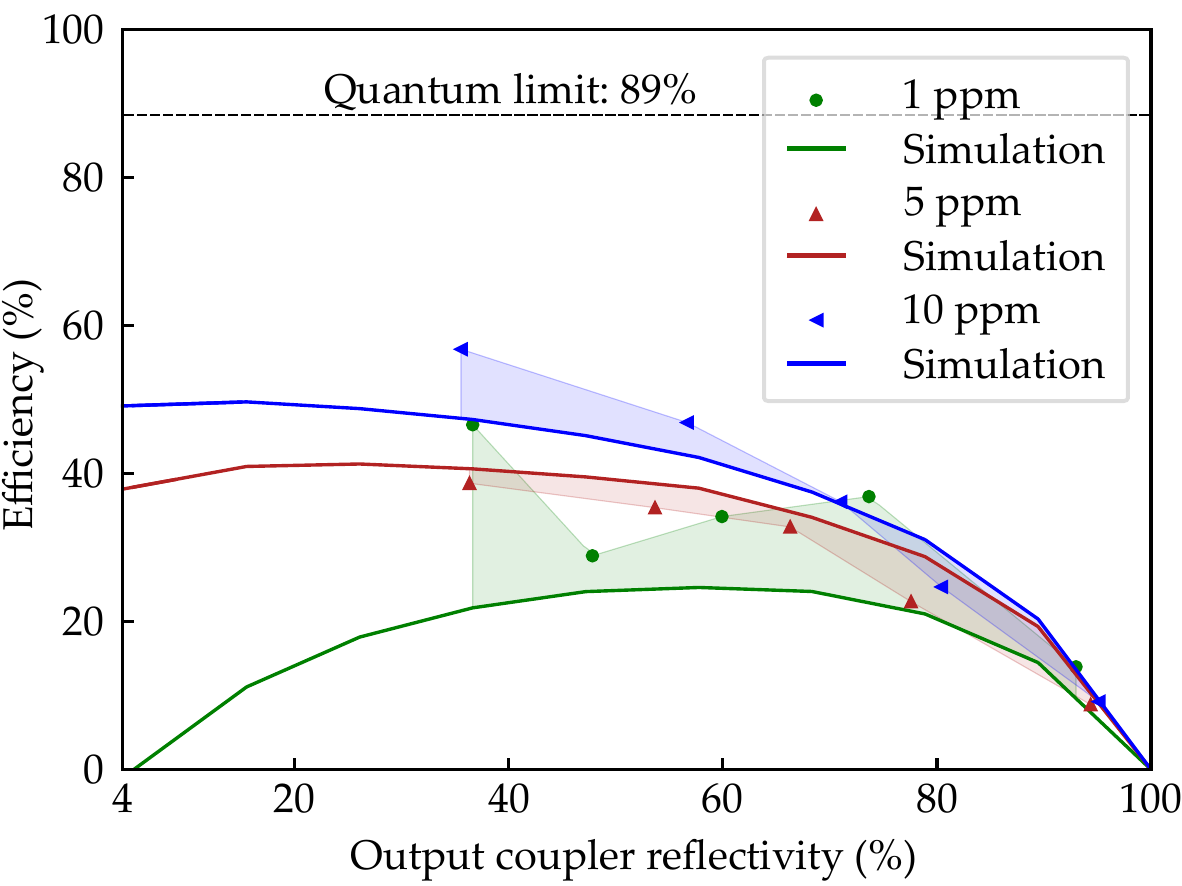}
	\caption{Comparison of the laser efficiencies of all configurations obtained by numerical (solid lines) and experimental results (markers).}
	\label{fig:slopes}
\end{figure}

As can be seen from Fig. \ref{fig:slopes}, where the lasing efficiencies of all configurations are plotted, our numerical model was able to fully support the experimental data of the $5\,ppm$ and $10\,ppm$ doped lasers (Fig. \ref{fig:slopes}). Somewhat higher deviations between numerical and experimental results were obtained for the $1\,ppm$ doped fiber. We assume an experimental uncertainty due to a nonuniform distribution of RB molecules in the RB:POF. To ensure that the results do not depend on a particular piece of fiber, the experimental results were confirmed with different pieces of the fiber. Understanding this behaviour of the RB(1\,ppm):POF laser needs further investigation.

\subsection{Long-Term Stability of the Laser Operation}
\begin{figure}[H]
	\centering
	\includegraphics[width=7.5 cm]{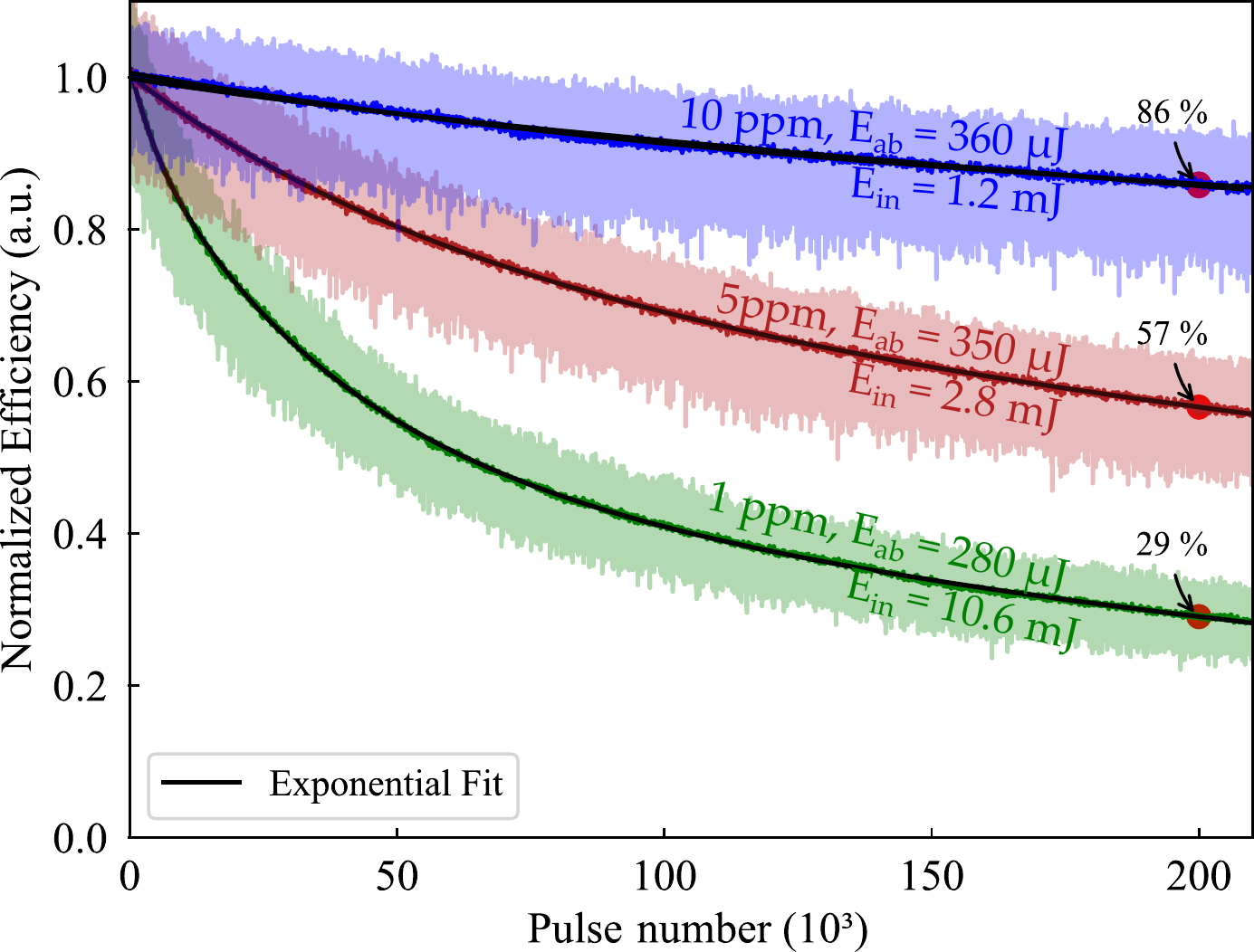}
	\caption{Long-term performance of the RB:POF laser with 1, 5 and 10\,ppm doping concentration. The pump energy was scaled towards comparable absorption within each fiber. The measurements are corrected for pump energy fluctuations. Exponential functions were fitted to the data revealing the $50\,\%$-lifetimes.}
	\label{fig:longterm}
\end{figure}

Bleaching is the well-known degradation mechanism which reduces the laser output energy of dye doped laser materials. Long-term measurement of the output energy was performed to characterize this degradation mechanism in our laser systems. The results are shown in Fig. \ref{fig:longterm}. The reflectivity of the OC was set to $70\,\%$, and the efficiency of lasers with doping concentrations of $1\,ppm$, $5\,ppm$ and $10\,ppm$ were recorded for $200\,k$ shots at $10\,Hz$ corresponding to 5 hours and 33 minutes with a photodiode and an analog/digital IO device connected to a computer via USB. The incident pump energy differed as the absorbed pump energy scales with the doping concentration, but all fibers had to absorb the same amount of energy for better comparability.
The normalized laser efficiency was calculated for better comparability and to prevent falsification due to pump fluctuations.  

The output energy of all lasers decreased significantly within the $200k$ shots. The normalized efficiency dropped to $86\,\%$, $57\,\%$, $29\,\%$ for the $10\,ppm$, $5\,ppm$ and $1\,ppm$ doped fiber, respectively. Fitting a mono-exponential function to the measured curves revealed the $50\,\%$-lifetime of the fiber lasers to be $905\,k$, $280\,k$ and $63\,k$ shots for the $10\,ppm$, $5\,ppm$ and $1\,ppm$ doped fiber, respectively. Compared with the value of $200\,k$ reported by Kuriki et al. in \cite{kuriki2}, this constitutes a 4.5-fold improvement of the $50\,\%$-lifetime. Note that our lasers were optimized towards high efficiencies. Optimization towards long-term stability could presumably lead to even higher $50\,\%$-lifetimes.

\subsection{Spectral Analysis}
\begin{figure}[H]
	\centering
	\includegraphics[width=15.5cm]{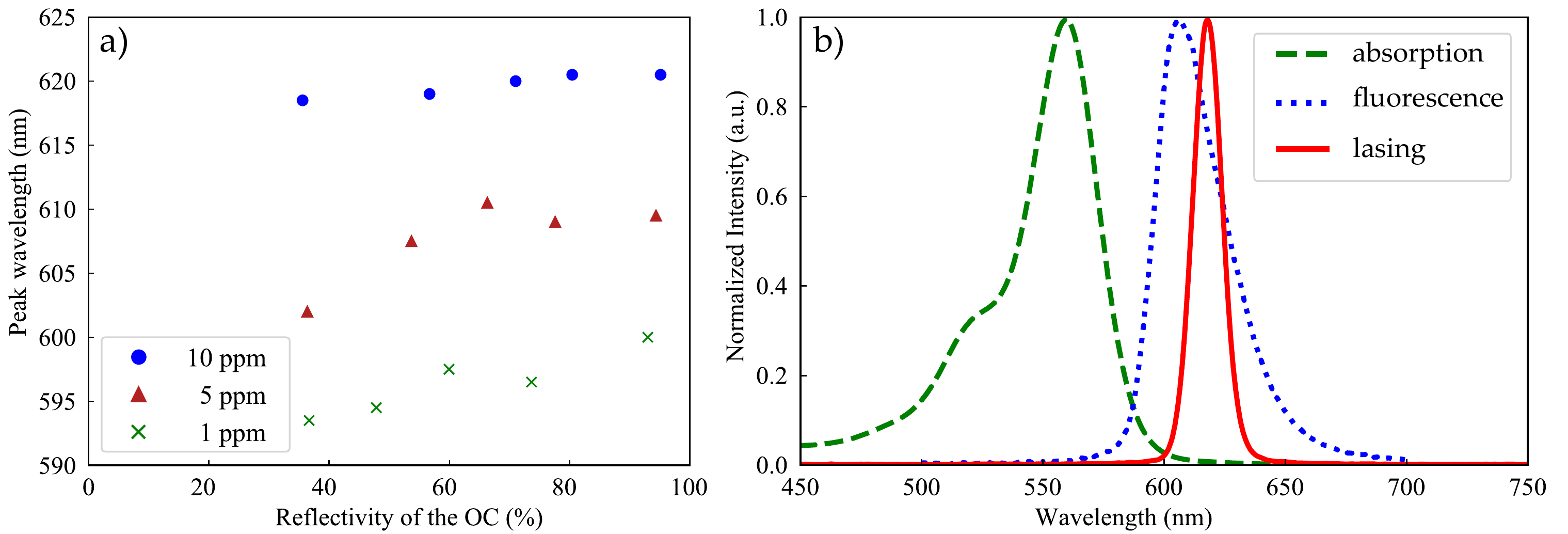}
	\caption{a) Representative absorption, fluorescence and laser spectrum highlight red shift of the emission wavelength during laser operation. b) Peak wavelength for different resonator configurations. The tuning range is approximately $28\,nm$ for the range of doping concentrations considered and $9\,nm$ within one doping concentration when changing the reflectivity of the output coupler.}
	\label{fig:speclaser}
\end{figure}

The output wavelength of the lasers changed depending on the doping concentration and the OC reflectivity, as it can be deduced from the plot in Fig. \ref{fig:speclaser} a). The wavelength shifted to longer wavelength for increasing doping concentrations resulting in a total wavelength shift of $\Delta \lambda \approx 28\,nm$. For decreasing OC reflectivities it shifted to shorter wavelength resulting in a wavelength shift of $\Delta \lambda \approx 9 \,nm$ within one doping concentration. The underlying mechanism is enabled by the spectral overlap of absorption and emission spectra. The laser wavelengths are thus spectrally shifted in comparison with spontaneous emission as it can be seen exemplarily in Fig. \ref{fig:speclaser} b).
Fibers with high doping concentrations show a higher number of RB molecules and thus reabsorption by an RB molecule is more likely than in fibers with low doping concentrations.
Analogously, a high OC reflectivity leads to more roundtrips of a photon within the resonator, resulting in a high probability for reabsorption. The reabsorbed photons can be reemitted which leads to the observed red shifts \cite{arrue2}. 
These results show one opportunity to tune the wavelength of our RB:POF lasers which can be used to extend the functionality and field of applications \cite{mhibik, sihna, peng}.

\subsection{Numerical Results on the Wavelength Tunability}

\begin{figure}[H]
	\centering
	\includegraphics[width=8cm]{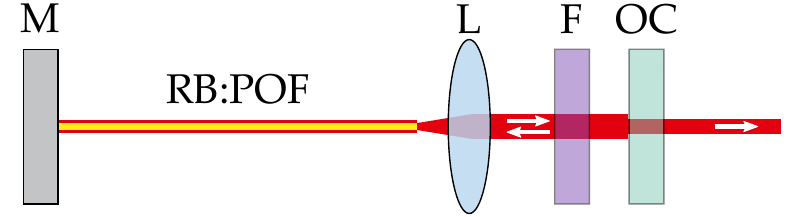}
	\caption{Schematic setup assumed for the numerical investigations on the tunability of a RB:POF laser. M: mirror; L: lens; F: spectral filter; OC: output coupler.}
	\label{fig:tunableLaser}
\end{figure}

As the tuning range observed with different resonator configuration was still much smaller than the fluorescence bandwidth, we applied our rate equation model to simulate the potential wavelength tunability of our RB:POFs (c.f. sec. \ref{sec:numerical}). In order to numerically treat this approach, a wavelength dependent output coupling loss was implemented representing the combination of a spectral filter with a bandwidth of $\Delta \lambda = 5\,nm $ and output coupling mirror as sketched in Fig. \ref{fig:tunableLaser}. Further input parameters were chosen according to the experimental values which are described in sec. \ref{sec:manufacturing} and \ref{sec:laser}.

\begin{figure}[H]
	\centering
	\includegraphics[width=16cm]{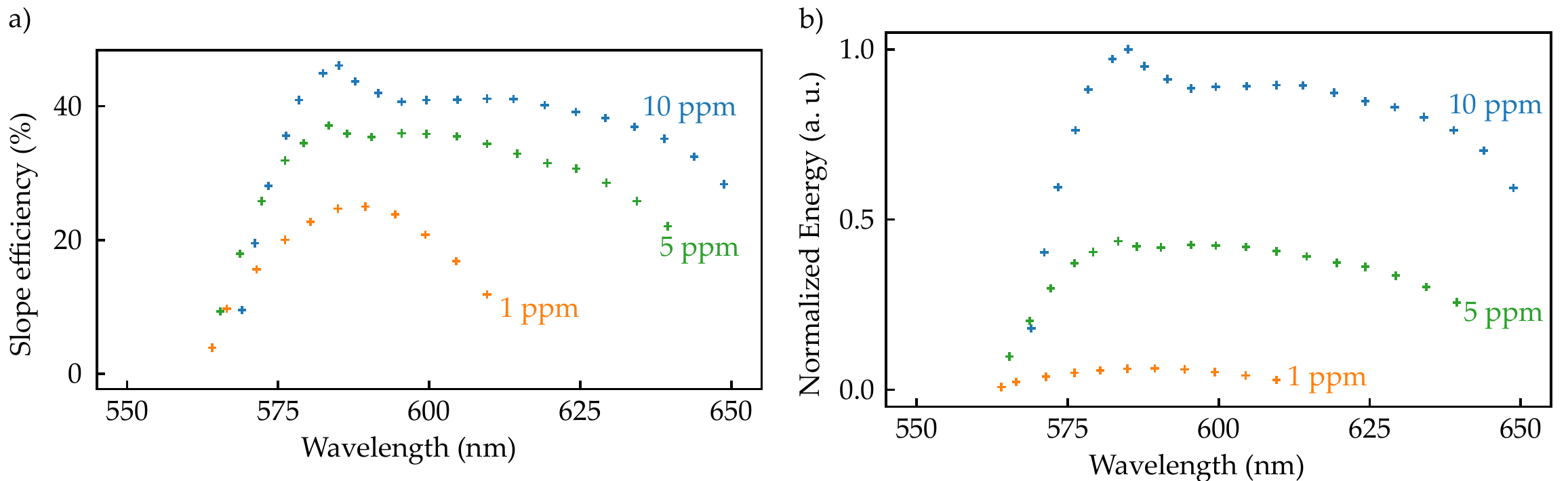}
	\caption{a) Simulated slope efficiency for a tunable polymer fiber laser consisting of our RB:POF. Efficiencies of more than $40\,\%$ are possible as well as a tuning range of more than $80\,nm$. b) Corresponding output energy of such tunable RB:POF laser. Highest output energies are expectable for $10\,ppm$ doped fibers.}
	\label{fig:tunableSim}
\end{figure}

The simulations yielded a tuning range of several $10\,nm$ as can be seen from fig. \ref{fig:tunableSim}, where the spectral dependence of the slope efficiency as well as the normalized output energy is shown. The slope efficiency and output energy show a maximum for $5\,ppm$ and $10\,ppm$ around $580\,nm$. This may be caused by the local minimum in the spectral attenuation in interplay with the specific spectral location of the emission (c.f. Fig. \ref{fig:spectral}). The lasers with all three doping concentrations show high slope efficiencies. RB:POFs doped with $10\,ppm$ can perform with a slope efficiency of more than $40\,\%$ according to our numerical results. Even a $1\,ppm$ doped RB:POF laser yields a slope efficiency of more than $20\,\%$. The tuning range of the lasers broadens with increasing doping concentrations from $> 40\,nm$ for $1\,ppm$ to $>80\,nm$ for $10\,ppm$ due to higher gain for higher doping concentrations. 

Exemplary, the RB($1\,ppm$):POF laser yields only about $10\,\%$ of output energy compared to a RB($10\,ppm$):POF laser. While the slope efficiencies for lasers with all three doping concentrations are satisfactory, output energies scale sensitively with the doping concentration in our simulations which is in accordance with experiments (c.f. sec. \ref{subsec:laseff}). To realize such tunable laser is possible in principle and will be adressed in the future to provide extension of the functionality and field of applications.

\section{Conclusions}

The route to high efficient dye doped POF lasers was presented in this paper and includes optimization of the pump geometry, pump wavelength, resonator quality, doping concentration and accompanying numerical treatment.
We presented polymer fiber lasers doped with RB which provide output energies up to $1.65\,mJ$, high slope efficiencies up to $56\,\%$, and $50\,\%$-lifetimes of more than $900\,k$ shots. These characteristics constitue a significant optimization compared to Rhodamine doped POF lasers reported previously \cite{kuriki1, kuriki2, al-shamiri, arrue3}. Additionally, coarse wavelength tunabilty of $28\,nm$ can be achieved by varying the doping concentration and fine tunabilty within $9\,nm$ is enabled through change of the resonator quality. 
The numerical modelling of the RB:POF lasers yielded accurate results compared to the experiments. Futhermore, simulations predict a broad spectral tuning range $>80\,nm$ including slope efficiencies of $>40\,\%$ by using a wavelength tunable RB:POF laser setup. Setting up such tunable RB:POF laser is aimed at in the future.

In the future, a work-around strategy is needed to replace the complex and expensive pump laser. Starting point for such further optimization might be the energy storage capacity which is described through the fluorescence lifetime of the laser dye. Prospectively, organic laser active media with cheaper adressable pump transitions such as electically pumped organic laser diodes will be of great interest to investigations on integrated photonic LoC devices. Further attempts to optimize the performance of polymer fiber lasers may be adressed to the attenuation of the material as well as the stability and fluorescence lifetime of the incorporated dye. Higher integration grades may be achieved by sputtering resonator mirrors to the fiber facets. Polymer fiber lasers and amplifiers can be applied to provide wavelength tunable, high efficient and energy delivering light sources at reasonable financial effort. The polymer basis of such sources leads to suitability and high integratebility for LoC-devices.

\vspace{6pt} 



\authorcontributions{conceptualization, S.U., S.S., and A.R.;
	methodology, S.S. and J.T.; 
	software, J.T.; 
	validation, S.U., J.T., S.S. and A.R.; 
	formal analysis, S.U. and S.S.; 
	investigation, S.U., J.T., F.J., J.K., P.Y.A. and S.S; 
	resources, J.N., H-H.J.; 
	data curation, S.U. and S.S.; 
	writing--original draft preparation, S.S., S.U., J.T.; 
	writing--review and editing, S.S., A.R., J.N., D.K. and D.R.; 
	visualization, S.U. and S.S.; 
	supervision, A.R., D.R.; 
	project administration, H.-H.J. and D.R.; 
	funding acquisition, W.K. and D.R.}

\funding{We gratefully acknowledge the state of Lower Saxony and the European Union for funding the LaPOF research network (EFRE-SER 2014-2020, 85003655 and 85003502). Collaborations of the involved institutions are also funded  by  the  Deutsche  Forschungsgemeinschaft  (DFG,  German  Research  Foundation)  under  Germany’s  Excellence Strategy within the Cluster of Excellence PhoenixD (EXC 2122, Project ID 390833453).}

\conflictsofinterest{The authors declare no conflict of interest. The funders had no role in the design of the study; in the collection, analyses, or interpretation of data; in the writing of the manuscript, or in the decision to publish the results.} 

\reftitle{References}

\end{document}